\documentclass[twocolumn,showpacs,showkeys,preprintnumbers,amsmath,amssymb,prl]{revtex4}

\usepackage{graphicx}
\usepackage{dcolumn}
\usepackage{bm}

\begin{document}


\title{Complex magnetic monopoles and geometric phases around diabolic and exceptional points}

\author{A. I. Nesterov}
   \email{nesterov@cencar.udg.mx}

\author{F. Aceves de la Cruz}%
 \email{fermin@udgphys.intranets.com}
\affiliation{Departamento de F{\'\i}sica, CUCEI, Universidad de Guadalajara,
Av. Revoluci\'on 1500, Guadalajara, CP 44420, Jalisco, M\'exico}

\date{\today}

\begin{abstract}
We study the geometric phase (GP)in presence of diabolic (DP) and exceptional
(EP) points. While the GP associated with the DP is the flux of the Dirac
monopole, the GP related to the EP, being complex one, is described by the flux
of complex magnetic monopole. For open systems, in week-coupling limit, the
leading environment-induced contribution to the real part of complex GP is
given by a quadrupole term, and to its imaginary part by a dipolelike field. We
find that the GP has a finite gap at the DP and infinite one at the EP.
\end{abstract}

\pacs{03.65.Vf, 14.80.Hv, 03.65.-w, 03.67.-a, 11.15.-q}

\keywords{Berry phase, Dirac monopole, complex geometric phase }

\maketitle

For a quantum-mechanical system in the commonest case of double degeneracy with
two linearly independent eigenvectors, the energy surfaces form the sheets of a
double cone, and its apex is called a ``diabolic point'' (DP) \cite{BW}. In the
context of Berry phase the DP is associated with `fictitious magnetic monopole'
located at the DP \cite{B0,BW}. In opposition to the DP, at which the
eigenvalues coincide while the eigenvectors still remain distinct, exceptional
points (EP) occur when the eigenvalues and eigenvectors coalesce. Since for
Hermitian operators the coalescence of eigenvalues results in different
eigenvectors, the EP degeneracy does not exists for the Hermitian operators and
is associated with open quantum systems and non-Hermitian physics.

While notion of the geometric phase (GP) for pure states is well defined, the
definition of a GP in open quantum systems is still an unsolved problem. The
first important step towards the consistent description of geometric phase for
open systems was done by Garrison and Wright \cite{GW}. They removed the
restriction to unitary evolution considering the quantum system governed by
non-Hermitian Hamiltonian. Uhlmann was the first to introduce the concept of GP
to the case of mixed states employing the density matrix approach
\cite{UH,UH_1}. However the physical interpretation of the Uhlmann GP for
density matrix is not clear. Recently, an alternative definition of the GP for
mixed states has been proposed by Sj\"oqvist {\em et al}  in the experimental
context of quantum interferometry \cite{SPEA}.

In this Letter we consider the GP in presence of DPs and EPs. We show that for
general non-Hermitian system the related GP is described by {\em complex
magnetic monopole}. We analyze GPs near the degeneracy and find that the GP is
nonanalytical and, having a finite gap at the DP, diverges at the EP.

{\em General results on behavior of the eigenvectors at diabolic and
exceptional points.--} It is known that in parameter space, a set of
exceptional points defines a smooth surface of codimension 2 for
symmetric/nonsymmetric complex matrix, codimension 1 for a real nonsymmetric
matrix, and EP's do not exist for real symmetric or Hermitian matrix
\cite{Arn}.

Let $H(X)$ be a complex $N\times N$ matrix smoothly dependent on $m$ real
parameters $X_i$ ($i$ runs from 1 to m).  For $\lambda_k(X)$ being the
eigenvalues of $H(X)$, we denote by $|\psi_k(X)\rangle$ and
$\langle\tilde\psi_k(X)|$ the corresponding right and left eigenvectors:
$H|\psi_k\rangle = \lambda_k|\psi_k\rangle, \quad \langle\tilde\psi_k|H=
\lambda_k\langle\tilde\psi_k|$. Both systems of left and right eigenvectors
form a bi-orthogonal basis \cite{MF}:
\begin{eqnarray}
\sum_k\frac{|\psi_k\rangle\langle\tilde\psi_k|}{\langle\tilde\psi_k|\psi_k\rangle}=
1 \quad \langle\tilde\psi_k|\psi_{k}\rangle = 0,\quad k \neq k' \label{Beq}.
\end{eqnarray}

Using the decomposition of unity (\ref{Beq}), one obtains $|\Psi\rangle =
\sum_i \alpha_i |\psi_k\rangle$ and $\langle\widetilde\Psi|= \sum_i \beta_i
\langle\tilde\psi_i|$, where
\begin{align}\label{Eq23}
\alpha_i =
\frac{\langle\tilde\psi_i|\Psi\rangle}{\langle\tilde\psi_i|\psi_{i}\rangle},
\quad \beta_i =
\frac{\langle\widetilde\Psi|\psi_{i}\rangle}{\langle\tilde\psi_i|\psi_{i}\rangle}.
\end{align}

We assume that EP occurs for some value of parameters $X=X_c$. At the EP the
eigenvalues, say $n$ and $n+1$, coalesce: $\lambda_n(X_c)=\lambda_{n+1}(X_c)$,
and the corresponding eigenvectors coincide yielding a single eigenvector
$|\psi_{\rm EP}\rangle =|\psi_n(X_c)\rangle=|\psi_{n+1}(X_c)\rangle$. Now,
applying (\ref{Beq}) for $k=n$ and $k=n+1$ we find that at the EP the
normalization condition is violated: $\langle\tilde\psi_{\rm EP}|\psi_{\rm
EP}\rangle =0$. This leads to the serious consequences for the global behavior
of the states in parameter space.

Since at the EP both eigenvalues and eigenvectors merge forming a Jordan block,
it is convenient to introduce the orthonormal basis of the related invariant
2-dimensional subspace as follows:
\begin{eqnarray}
\langle n|n\rangle =1, \quad \langle n+1|n+1\rangle =1, \quad \langle
n|n+1\rangle =0
\end{eqnarray}
Assuming that all other eigenstates are non-degenerate, we find that the set of
vectors $\{  \langle\tilde\chi_k |, |\chi_k \rangle\}$ , such that for
$k=n,n+1$ one has $ |\chi_k \rangle = |k\rangle,\;  \langle\tilde\chi_{k} |=
\langle k|$, and
\begin{eqnarray*}
|\chi_k \rangle = \frac{|\psi_k\rangle}{\sqrt{\langle
\tilde\psi_k|\psi_k\rangle}}, \quad \langle\tilde\chi_k | =
\frac{\langle\tilde\psi_k |}{\sqrt{\langle
 \tilde\psi_k|\psi_k\rangle}}, \; {\rm for}\; k\neq n,n+1,
\end{eqnarray*}
forms the bi-orthonormal basis. Using this basis we expand an arbitrary vector
as $|\psi\rangle =  \sum c_k (X)|\chi_k (X) \rangle$, with the coefficients of
expansion being $c_k = \langle \tilde\chi_k|\psi\rangle $.

From the orthogonality condition, one can see that if $|\psi(X)\rangle
\rightarrow |\psi_{EP}\rangle$ while $X \rightarrow X_c$, then all $c_k$
($k\neq n,n+1$) vanish at EP. Thus in the neighborhood of EP only the terms
related to the invariant subspace make substantial contributions and the
$N$-dimensional problem becomes effectively two-dimensional \cite{Arn,KMS}.
Similar conclusion is valid for the DP's, excepting that at the DP the
eigenvectors don't coalesce. The detailed study of the associated
two-dimensional problem will be presented in the following sections.

{\em Geometric phase for non-Hermitian systems.--} The GP for non-Hermitian
systems were studied by various authors (see, for instance,
\cite{GW,B,B1,B2,H1,H2,KKm} and references therein). Following \cite{GW}, let
us consider the time dependent Schr\"odinger equation and its adjoint equation:
\begin{align}\label{S1}
i\frac{\partial }{\partial t}|\Psi(t)\rangle = H(X(t))|\Psi(t)\rangle, \\
-i\frac{\partial }{\partial t}\langle\widetilde\Psi(t)| =
\langle\widetilde\Psi(t)|H(X(t))\label{S2}.
\end{align}

Let $\langle\tilde\psi_n(X)|$ and $|\psi_n(X)\rangle$ be left (right)
eigenstates corresponding to the eigenvalue $E_n$, then in adiabatic
approximation the complex geometric phase is given by the integral
\cite{GW,B1,B2}
\begin{align*}
\gamma_n = i\oint_C
\frac{\langle\tilde\psi_n(X)|d\psi_n(X)\rangle}{\langle\tilde\psi_n(X)|\psi_n(X)\rangle}=
i\oint_C \langle\tilde\chi_n(X)|d\chi_n(X)\rangle.
\end{align*}
Validity of the adiabatic approximation is defined by the following condition
\cite{GW}:
\begin{equation}
\label{Eq24}
    \sum_{m\neq n}\bigg|\frac{\langle\tilde\psi_m|\partial H/\partial
    t|\psi_n(X)\rangle}{(E_m - E_n)^2}\bigg|\ll 1
\end{equation}
This restriction is violated nearby the degeneracies related to any of  DP or
EP, where the eigenvalues coalesce.

Since the adiabatic approach cannot be applied in the neighborhood of
degeneracy, further we will consider non adiabatic generalization of Berry's
phase introduced by Aharonov and Anandan \cite{AA} and extended by Garrison and
Wright \cite{GW} to the non-Hermitian systems as follows. Let an adjoint pair
$\{|\Psi(t)\rangle, \langle\widetilde\Psi(t)|\}$ being a solution of Eqs.
(\ref{S1}), (\ref{S2}) satisfies the following condition:
\begin{align*}
|\Psi(T)\rangle = \exp(i\varphi)|\Psi(0)\rangle,\;  \langle\widetilde\Psi(T)|=
\exp(-i\varphi)\langle\widetilde\Psi(0)|,
\end{align*}
where $\varphi$ is a complex phase, and $\{|\chi(t)\rangle,
\langle\tilde\chi(t)|\}$ is a modified adjoint pair such that
\begin{align*}
|\chi(t)\rangle = \exp(-if(t))|\Psi(t)\rangle, \; \langle\tilde\chi(t)|=
\exp(if(t))\langle\widetilde\Psi(t)|,
\end{align*}
where $f(t)$ is any function satisfying $f(T) - f(0)= \varphi(0)$. The total
phase $\varphi$ calculated for the time interval $(0,T)$ may be written as
$\varphi = \gamma + \delta$, where the geometric phase $\gamma$ is given by
\begin{equation}\label{Eq27}
 \gamma = i\int_0^T \langle\tilde\chi(t)|\frac{\partial}{\partial
 t}\chi(t)\rangle dt
\end{equation}
and for the ``dynamical phase'' one has
\begin{equation}\label{Eq27a}
 \delta = -\int_0^T \langle\tilde\chi(t)|H|\chi(t)\rangle dt.
\end{equation}
This yields the connection one-form and the curvature two-form as follows:
\begin{equation}\label{Eq11}
A= i\langle\tilde\chi|d\chi\rangle, \quad F=dA.
\end{equation}

{\em Two-level system.--} Let us consider a two-level system described by
generic non-Hermitian Hamiltonian:
\begin{equation}\label{eqH2a}
H=\left(
  \begin{array}{cc}
    \lambda_0 + Z & X-iY \\
   X+iY & \lambda_0 - Z\\
  \end{array}
\right), \quad X,Y,Z \in \mathbb C
 \end{equation}
Further we consider a complex matrix $H$ as depending on three complex
parameters $X,Y$ and $Z$. While the DP is just a point in 3-dimensional complex
space $\mathbb{C}^3$, a set of EP being a hypersurface of complex codimension 1
in $\mathbb C^3$, defines a smooth surface of codimension 2 in 6-dimensional
real space \cite{Arn,KMS}.

The solution of the eigenvalue problem
\begin{equation}\label{EP1}
H|u\rangle =\lambda|u\rangle, \quad \langle \tilde u| H =\lambda \langle \tilde
u|,
\end{equation}
where $|u\rangle$ and $\langle \tilde u|$ are the right and left eigenvectors,
respectively, is given by $\lambda_{\pm} = \lambda_0 \pm \sqrt{X^2 +Y^2 +
Z^2}.$ The left and right eigenvectors are found to be
\begin{align}
&|u_{-}\rangle = \left(\begin{array}{c}
-e^{-i\varphi}\sin\frac{\theta}{2}\\
\cos \frac{\theta}{2} \end{array}\right), \; \langle \tilde u_{-}|
=\bigg(-e^{i\varphi}\sin\frac{\theta}{2}, \cos\frac{\theta}{2}\bigg)
\nonumber \\
&|u_{+}\rangle = \left(\begin{array}{c}
                  e^{-i\varphi}\cos\frac{\theta}{2}\\
                  \sin\frac{\theta}{2}
                  \end{array}\right),
\langle \tilde u_{+}| = \bigg(e^{i\varphi}\cos\frac{\theta}{2},
\sin\frac{\theta}{2}\bigg)  \label{r}
\end{align}
where
\begin{align}\label{Eq15}
&&\cos\frac{\theta}{2}= \sqrt{\frac{R+Z}{2R}}, & \quad
\sin\frac{\theta}{2}=\sqrt{\frac{R-Z}{2R}}, \\
&&e^{i\varphi} = \frac{X + iY}{\sqrt{R^2 -Z^2}}&,\quad R= \sqrt{X^2 +Y^2 +Z^2}.
\end{align}
and $\theta, \varphi$ are the complex angles of the complex spherical
coordinates $(R,\theta, \varphi)$. For $R\neq 0$ the following relationships
hold: $ \langle \tilde u_{\pm}|u_{\mp}\rangle = 0, \; \langle \tilde
u_{\pm}|u_{\pm}\rangle = 1$.

As can be easily seen, the coupling of eigenvalues occurs when $X^2 +Y^2 + Z^2
=0$. There are two cases: $\theta=0, \; \varphi =0$ and $\theta= i\infty,\;
\varphi =0 $. The first case yields two linearly independent eigenvectors and
the point of coupling is known as the diabolic point (DP). Since for $\theta
\rightarrow i \infty$ the eigenvectors merge at the coupling point, the second
case corresponds to the EP. At the DP we obtain
\begin{align}
&|u_{+}\rangle = \left(\begin{array}{r}
                  1\\
                  0
                  \end{array}\right),&
|u_{-}\rangle = \left(\begin{array}{c}
0\\
1 \end{array}\right) \label{right}\\
&\langle \tilde u_{+}| = (1, 0), &\langle \tilde u_{-}| =(0, 1) \label{left},
\end{align}
and at the EP
\begin{eqnarray}
|u_{+}\rangle = |u_{-}\rangle \propto \left(\begin{array}{c}
                  e^{-i\varphi}\\
                  i
                  \end{array}\right), \quad
\langle \tilde u_{+}| = \langle \tilde u_{-}| \propto ( e^{i\varphi},i).
 \label{theta4}
\end{eqnarray}
This implies the violation of the normalization at the EP, and one has $\langle
\tilde u_{\pm}|u_{\pm}\rangle = 0$.

Applying (\ref{Eq11}) to $|u_{\pm}\rangle$ and  $\langle \tilde u_{\pm}|$, we
obtain
\begin{align}
\label{Eq12} A^{\pm}= \mp q(1 \pm\cos\theta)d\varphi, \quad F^{\pm}=
q\sin\theta\; d\theta \wedge d\varphi
\end{align}
where $q=1/2$. This describes the complex ``magnetic monopole" with a charge
$q$ and the field $\mathbf B$ given by
\begin{equation}\label{Eq4a}
\mathbf B = q\frac{\mathbf R}{R^3}
\end{equation}
where  $\mathbf R = (X,Y,Z)$ and $R=\sqrt{ X^2+ Y^2 +Z^2}$. As can be easily
seen the field of the monopole can be written as $B^i =
-\partial\varPhi/\partial X^i$, where the potential $\varPhi = q/R$.

Computation of geometric phase yields $\gamma = \oint_{\mathcal C} A$,
integration being performed over the contour $\mathcal C$ on the complex sphere
$S^2_c$. Applying the Stokes theorem we obtain
\begin{equation}\label{GP2}
\gamma = \int_{\Sigma} F = q\Omega(\mathcal C)
\end{equation}
where $\Sigma$ is a closed surface with the boundary $\mathcal C= \partial \Sigma $,
and $\Omega(\mathcal C)$ is the complex solid angle subtended
by the contour $\mathcal C$.

For DP formula (\ref{Eq4a}) reproduces the classical result by Berry on
two-fold degeneracy in parameter space \cite{B0}. For EP the field of the
corresponding ``monopole"  represents a complicated topological charge rather
than a pointlike magnetic charge.

\begin{figure}[tbh]
\begin{minipage}[]{8.5cm}
\scalebox{0.4}{\includegraphics{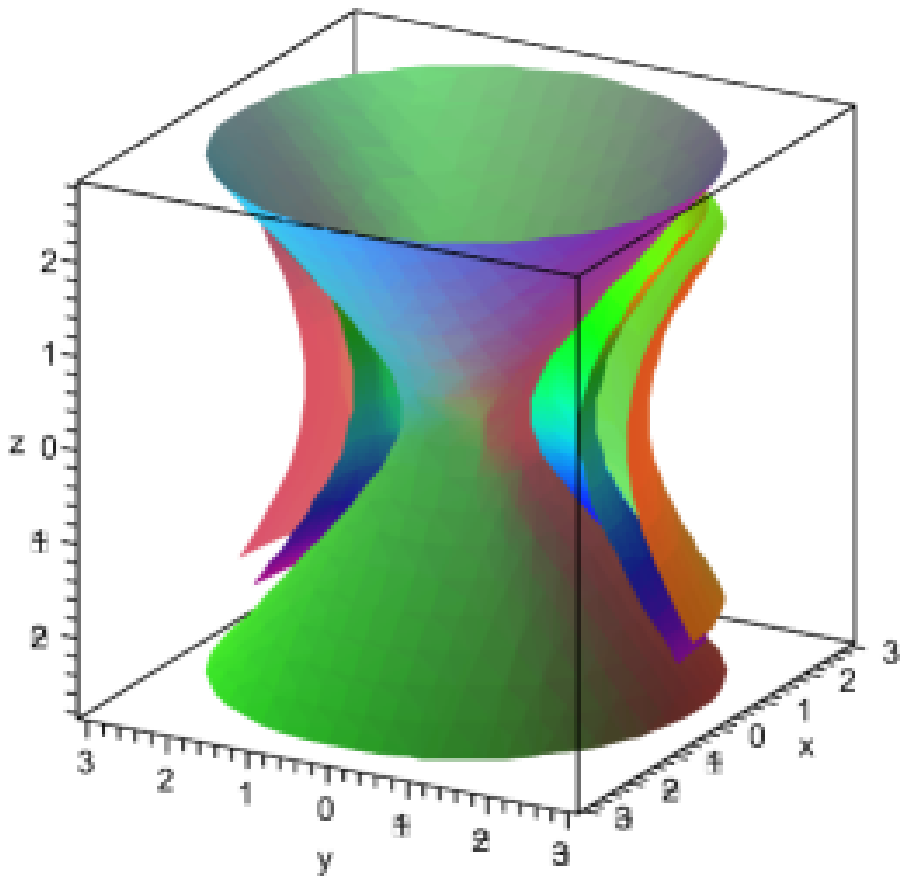}} \scalebox{0.2}{\includegraphics{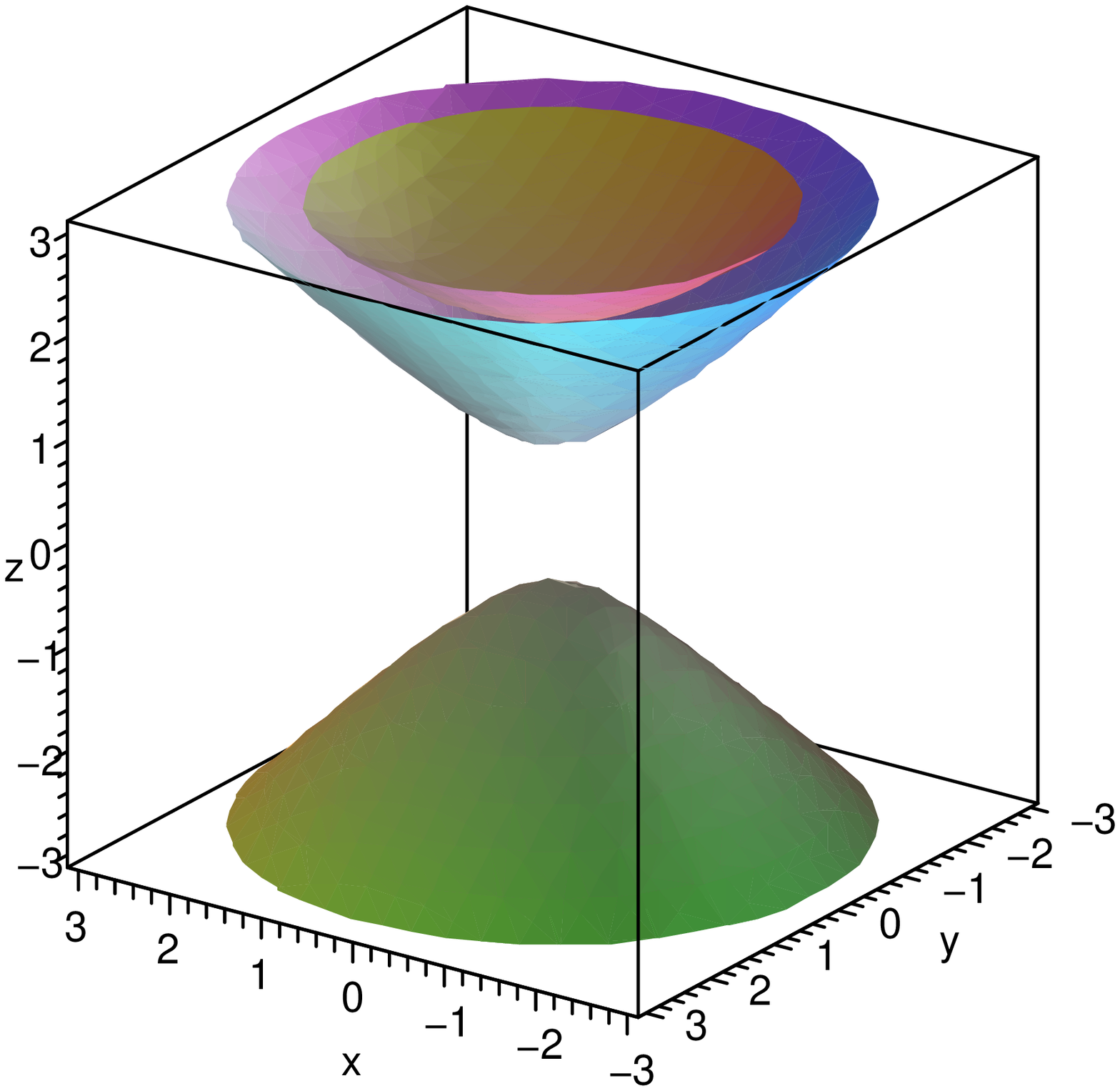}}
\caption{Imaginary hyperbolic monopole: the surfaces $\Im \varPhi = \rm const$
are presented (left). Real hyperbolic monopole: the surfaces $\Re \varPhi = \rm
const$ are plotted (right).\label{HMC}}
\end{minipage}
\end{figure}

{\em Hyperbolic monopole.--} Let us consider the following non-Hermitian
Hamiltonian:
\begin{align}\label{Eq17g}
H=\left(
  \begin{array}{cc}
    \lambda_0+ iz & x-iy \\
    x+iy & \lambda_0- iz \\
  \end{array}
\right), \quad x,y,z \in \mathbb R
\end{align}
The computation yields
\begin{equation}\label{Eq4}
\mathbf B =q\frac{\mathbf R}{R^3}, \quad \mathbf R = (x,y,iz)
\end{equation}
where $R=\sqrt{ x^2+ y^2 -z^2}$, and $\mathbf B$ is the field of ``monopole''
be called the {\em hyperbolic monopole} (Fig.\ref{HMC}).

The EP defined as the solution of the equation $R=0$, is represented by the
double cone with the apex at the origin of coordinates, and the DP is just
located at the origin of coordinates. For $R>0$ we obtain the imaginary
hyperbolic monopole ($\Re B=0$):
\begin{align}
A^{\pm}= \mp q(1 \mp \sinh\theta)d\varphi,\quad F = q\cosh\theta\; d\theta
\wedge d\varphi, \label{Eq20}
\end{align}
and for $R<0$ one has the real hyperbolic monopole ($\Im B=0$)(see
Fig.\ref{HMC}):
\begin{align}
A^{\pm}= \mp q(1 \mp \cosh\theta)d\varphi,\quad F = q\sinh\theta\; d\theta \wedge
d\varphi. \label{Eq20a}
\end{align}
{\em Complex Dirac monopole.--} For the complex Hamiltonian
\begin{align}\label{Eq17a}
H=\left(
  \begin{array}{ll}
    \lambda_0+ z -i\varepsilon & x-iy \\
    x+iy & \lambda_0- z +i\varepsilon \\
  \end{array}
\right), \quad x,y,z \in \mathbb R
\end{align}
the computation yields the field  $\mathbf B$ as follows:
\begin{equation}\label{Eq4g}
\mathbf B =\frac{q\mathbf R}{R^3}, \quad \mathbf R = (x,y,z-i\varepsilon)
\end{equation}
where $R=\sqrt{ x^2+ y^2 +z^2 -\varepsilon^2 - 2i\varepsilon z}$. For
$\varepsilon=0$ the obtained {\em complex Dirac monopole} becomes well-known
pointlike Dirac monopole.

\begin{figure}[tbh]
\begin{minipage}[]{8.5cm}
\scalebox{0.4}{\includegraphics{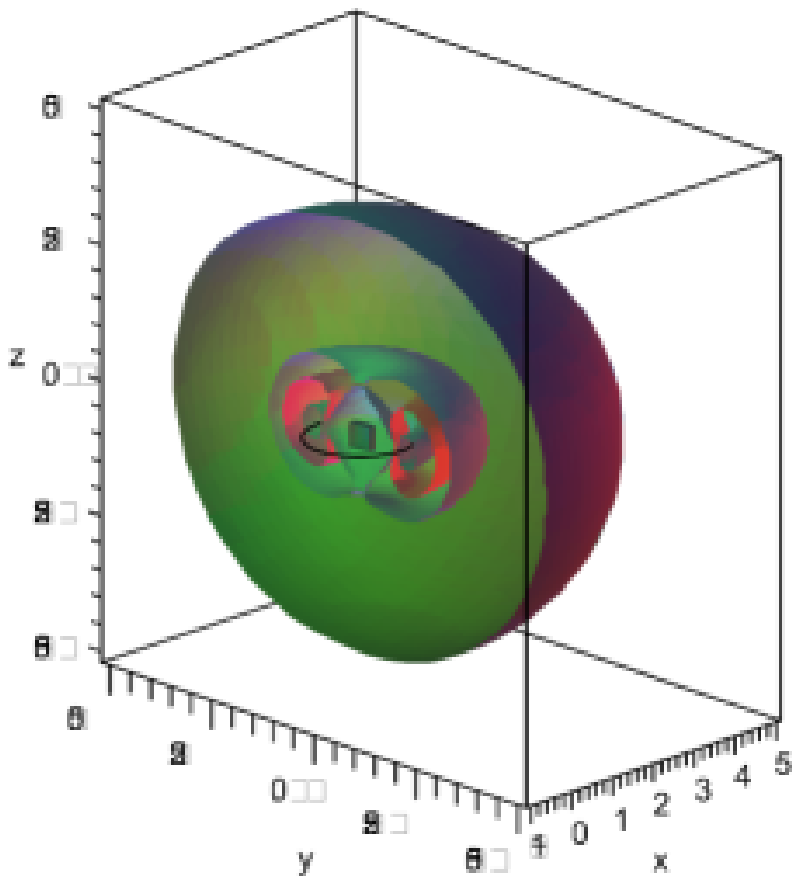}}
\scalebox{0.45}{\includegraphics{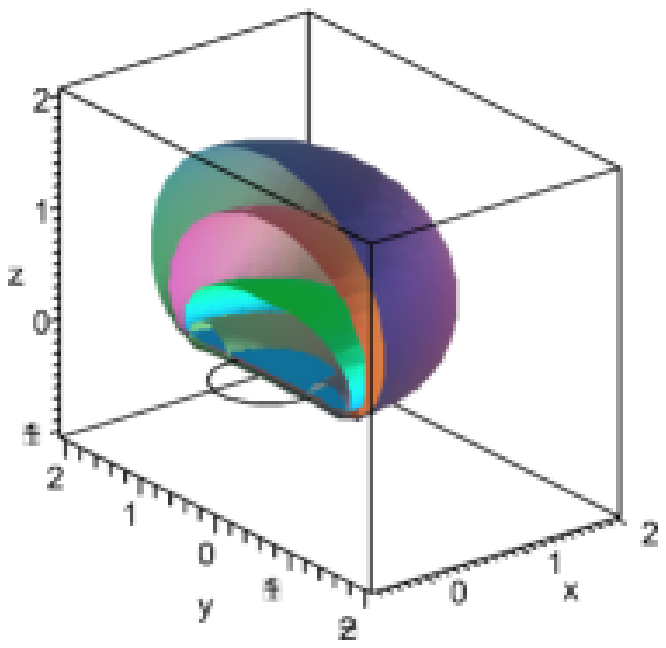}} \caption{Complex Dirac monopole. The
surfaces $\Re \varPhi = \rm const$ (left) and $\Im \varPhi = \rm const$ (right)
are plotted. The EP is appeared as the circle of the radius $r=1$ at the plane
$z=0$ \label{HME}}
\end{minipage}
\end{figure}

The EP being determined as the solution of equation $ x^2+ y^2 +z^2
-\varepsilon^2 + 2i\varepsilon z=0$ is the circle of the radius $\varepsilon$ a
the plane $z=0$. For $\varepsilon =1$ the  real and imaginary parts of
$\varPhi=\rm const$ are presented in Fig.\ref{HME}.

Let us write $R$ as follows: $R= \sqrt{r^2 - 2i\varepsilon r
\cos\chi-\varepsilon^2 }$, where $r= \sqrt{ x^2+ y^2 +z^2}$. We may expand
$1/R$ as
\begin{align}\label{Eq28}
\frac{1}{\sqrt{r^2 - 2i\varepsilon r\cos\chi-\varepsilon^2 }} =
\sum_{l=0}^{\infty}\frac{(i\varepsilon)^l}{r^{l+1}}P_l(\cos\chi),
\end{align}
$P_l(\cos\chi)$ being the Legendre polynomials. For $r \gg \varepsilon$, this
yields the following expansion of $\varPhi$:
\begin{align}\label{Eq29b}
\varPhi = \frac{q}{r} + i\frac{p \cos\chi}{r^2}- \frac{Q(3\cos^2\chi -1)}{2r^3}
+ \dots
\end{align}
where $q$ is the total charge of the monopole, $p=q\varepsilon$ is the dipole
moment, and $Q=q\varepsilon^2$ the quadrupole moment.

In the spherical coordinates $(r,\chi,\varphi)$ the GP for the
state $\lambda_{-}= \lambda_0 - R$ is given by
\begin{align}\label{Eq29a}
\gamma = q\oint_{\mathcal C}\bigg(1- \frac{r\cos\chi-i\varepsilon}{\sqrt{r^2-
2i\varepsilon r \cos\chi-\varepsilon^2}}\bigg)d\varphi
\end{align}
Using multipole expansion (\ref{Eq29b}), we obtain for $\gamma$ the following
expression:
\begin{align*}
\gamma = \gamma_M +i \oint_{\mathcal C} \frac{p \sin^2\chi}{r} d\varphi -
\oint_{\mathcal C} \frac{3Q \sin^2\chi \cos\chi}{2r^2} d\varphi+\dots .
\end{align*}
where $\gamma_M= q\oint_{\mathcal C}(1-\cos\chi)d\varphi$ is the contribution
of the Dirac pointlike monopole, second term describes the dipole contribution
to the imaginary part of the geometric phase and third term the quadrupole
contribution to its real part.

A complex Dirac monopole appears in wide class of open systems, where the
Hamiltonian
$$\tilde H= \mathbf B(t)\cdot \boldsymbol \sigma-
\frac{i}{2}\Gamma^\dagger \Gamma
$$
includes spontaneous decay $\Gamma = \sqrt{\varepsilon}\sigma_{-}$ as a source
of decoherence. For instance, it emerges in a two-level atom driven by periodic
electromagnetic field $E(t)= \Re (\mathcal E(t)\exp(i\nu t))$, with $\mathcal
E(t)$ being slowly varied, as follows. In the rotating wave approximation the
Schr\"odinger equation reads \cite{GW,LSS}
\begin{align}\label{Sch1}
\left(
  \begin{array}{c}
   \dot u_1 \\
   \dot u_2 \\
  \end{array}
\right) =
\left(
  \begin{array}{cc}
    -\frac{i}{2}\gamma_a & V^\ast e^{i\Delta t}\\
    V e^{-i\Delta t} & -\frac{i}{2}\gamma_b \\
  \end{array}
  \right)
\left(
  \begin{array}{c}
    u_1 \\
    u_2 \\
  \end{array}
\right)
\end{align}
where $\gamma_a, \;\gamma_b$ are decay rates for upper and lower levels
respectively, $\Delta= \omega - \nu$, $\omega= (E_a-E_b)$,
$V=\boldsymbol\mu^\ast \cdot \boldsymbol{\mathcal E}$, and $\boldsymbol \mu$ is
the electric dipole moment. Removing the explicit time dependence of the
Hamiltonian with the non unitary transformation \cite{GW}, one obtains $\tilde
H$ in the form of (\ref{Eq17a}). To compare our results with that found in
\cite{GW} we set $x= \Re V(t), \; y= \Im V(t), \; z= \Delta/2$ and
$\varepsilon= \delta=(\gamma_a- \gamma_b)/2$. Let us consider the closed curve
$\mathcal C$ parameterized by $\varphi$ with the complex angle
$\theta=\theta_0$. Then the GP of Eq.(\ref{Eq29a}) becoming
\begin{equation}\label{Eq30}
\gamma= \pi\bigg( 1- \frac{\Delta - i\delta}{\sqrt{|2V_0|^2+ (\Delta -
i\delta)^2}}\bigg)
\end{equation}
is the GP obtained by Garrison and Wright \cite{GW}.

Of the particular interest is the behavior of GP near the EP. For the resonance
frequencies ($\Delta=0$) we obtain
\begin{align}
 \Re \gamma= \Bigg \{
\begin{array}{l}
 \pi, \;{\rm if}\; r > \delta \\
  \pi \Big(1\pm \displaystyle\frac{\delta}{\sqrt{\delta^2 - r^2 }}\Big), \; {\rm if} \;
  r < \delta,
\end{array}
\end{align}
where the upper/lower sign corresponds to $\Delta \rightarrow \pm 0$ and we set
$r=|2V_0|$. This yields the singularity of GP at the EP ($r \rightarrow
\delta$), and the GP has a finite gap at the DP \cite{NF}.

In summary, we show that, while the DP is associated with Dirac magnetic
monopole, the EP is related to the complex magnetic monopole. We found that for
real part of GP the first correction to the flux of the Dirac monopole field is
given by the quadrupole term, and the expansion for its imaginary part starts
with the dipolelike field. Similar conclusion has been obtained for a two-level
spin-half system in a slowly varying magnetic field and weakly coupled to a
dissipative environment \cite{WMSG}. Note, our results are generic and should
be applied to any dissipative system with an accidental energy levels crossing.

{\em Concerning experiment.--} The singular behavior of the GP near the EP
could be observed in the experiments to measure mixed-states GP with neutrons o
with polarized light, using a Mach-Zhender interferometer \cite{BH,EAB}.

The authors thank A. B. Klimov, J. L. Romero and S. G. Ovchinnikov for helpful
discussions and comments. This work is supported by research grants SEP-PROMEP
103.5/04/1911 and CONACyT U45704-F.

\bibliography{dep_prl}

\end{document}